# Terrestrial Effects of Nearby Supernovae: Updated Modeling

Brian C. Thomas[1] and Alexander M. Yelland[1,2]
[1] Department of Physics and Astronomy, Washburn University, Topeka, KS 66621 USA; brian.thomas@washburn.edu
[2] Department of Physics, Massachusetts Institute of Technology, Cambridge, MA 02139, USA

## Abstract

We have reevaluated recent studies of the effects on Earth by cosmic rays (CRs) from nearby supernovae (SNe) at 100 and 50 pc, in the diffusive transport CR case, here including an early-time suppression at lower CR energies neglected in the previous works. Inclusion of this suppression leads to lower overall CR fluxes at early times, lower atmospheric ionization, smaller resulting ozone depletion, and lower sea-level muon radiation doses. Differences in the atmospheric impacts are most pronounced for the 100 pc case with less significant differences in the 50 pc case. We find a greater discrepancy in the modeled sea-level muon radiation dose, with significantly smaller dose values in the 50 pc case; our results indicate it is unlikely that muon radiation is a significant threat to the biosphere for SNe beyond 20 pc, for the diffusive transport case. We have also performed new modeling of the effects of SN CRs at 20 and 10 pc. Overall, our results indicate that, considering only the effects of CRs, the "lethal" SN distance should be closer to 20 pc rather than the typically quoted 8–10 pc. Recent work on extended SN X-ray emission indicates significant effects out to 50 pc and therefore the case is now strong for increasing the standard SN lethal distance to at least 20 pc. This has implications for studies of the history of life on Earth as well as considerations of habitability in the Galaxy.

*Unified Astronomy Thesaurus concepts:* Supernovae (1668); Ozone layer (1194)

## 1. Introduction

In recent years, multiple discoveries of live $^{60}$Fe on Earth (Knie et al. 2004; Ludwig et al. 2016; Wallner et al. 2016, 2020), the Moon (Fimiani et al. 2016), and in near-Earth space (Binns et al. 2016) have established the occurrence of at least one, and most likely several, core-collapse supernovae (SNe) within 50–100 pc from Earth (Breitschwerdt et al. 2016; Fry et al. 2016) around 2.6 million years ago. Possible effects on Earth's atmosphere, biosphere, and climate by such SNe have been investigated (Thomas et al. 2016; Melott et al. 2017; Thomas 2018; Melott et al. 2019; Melott & Thomas 2019; Melott et al. 2020; Thomas & Ratterman 2020). Predicted effects include depletion of stratospheric ozone with a subsequent surface-level increase in solar ultraviolet radiation; increased muon and neutron radiation exposure at Earth's surface and into oceans; a possible increase in wildfires following increased lightning rates; and minor climate changes due to changes in atmospheric ozone profiles.

Each of these recent studies relies on modeling the flux at the Earth of cosmic-ray (CR) protons accelerated by nearby SNe (hereafter, SNCRs). For certain cases in previous work diffusive transport of CRs was assumed; however, the calculations reported therein (Thomas et al. 2016; Melott et al. 2017) neglected to include a factor in the diffusive transport equation that results in a suppression of lower-energy CRs at early times. This suppression is due to the slower transport of lower-energy CRs through the interstellar magnetic field environment and is more important at greater distances. In this work we revisit that modeling with revised calculations and update the predicted effects in the 50 and 100 pc cases previously considered based on new results for the SNCR flux. We also extend our modeling to SNe at 20 and 10 pc.

Comparing to past work, here we find in general lower SNCR fluxes, especially at lower primary energies and earlier times, with subsequently smaller impacts on ozone and the sea-level radiation dose. Differences in ozone depletion are relatively small, but the muon dose is significantly changed. It should be noted that the comparisons here apply only to the diffusive CR transport cases in those earlier works; other cases presented there are unaffected.

## 2. Calculation of the SNCR Flux

Thomas et al. (2016) and Melott et al. (2017) describe the methods and results of calculations of the flux at Earth of CRs accelerated by SNe at 100 pc and 50 pc, respectively. The methods applied, and results, depend on assumptions regarding the interstellar medium between Earth and the SNe, most importantly the magnetic field environment. Since CRs (mostly protons) are charged particles, their path from the SN to Earth is strongly affected by magnetic fields in the intervening medium, modulated by the particle energy (less effect for higher kinetic energy). In Thomas et al. (2016) three cases were considered. Cases A and B model the propagation of CRs from an SN at 300 and 50 pc from a Galactic magnetic field line that passes through the solar system; Case B also assumes a low-energy cutoff of the flux, while Case A does not include such a cutoff. Case C considers an SN at 100 pc and assumes the solar system exists in a region of interstellar medium already excavated by earlier SN events, with the magnetic field environment dominated by a turbulent component of 0.1 μG. For that case, SNCR propagation can be modeled using the diffusion approximation. In Melott et al. (2017) two cases were considered. Case A assumes a Galactic magnetic field line that runs directly from the SN through the solar system. Case B in Melott et al. (2017) is equivalent to Case C in Thomas et al.

(2016), with only a turbulent magnetic field component, but for an SN placed at 50 pc instead of 100 pc.

In this work we use the magnetic field environment assumptions of Case C in Thomas et al. (2016) (Case B in Melott et al. 2017). In those previous works an early-time, low-energy suppression was not applied, leading to an overestimate of the CR flux in the diffusive transport cases. Here, we recompute the SNCR flux for SNe at 100 and 50 pc, and also perform new calculations for SNe at 20 and 10 pc.

We assume a purely turbulent magnetic field in the intervening space between the SNe and Earth, with field strength 0.1 $\mu$G (de Avillez & de Breitschwerdt 2005). The CRs propagate isotropically and we compute the flux at Earth using a diffusive approximation with $D(E) = D_0(E/E_0)^{1/3}$, $D_0 = 2 \times 10^{28}$ cm$^2$ s$^{-1}$, and $E_0 = 1$ GeV. Note that in Melott et al. (2017) the $E_0$ value is given as 10 GeV, which is a typographical error.

Given source spectrum $Q(E) = Q_0(E/E_0)^{-2.2} e^{-E/E_c}$ with $Q_0 = 3 \times 10^{52}$ protons GeV$^{-1}$ and $E_c = 1$ PeV, the CR density at distance $r$ and time $t$ after the initial release of CRs from the SN is given by $n(E, r, t) = \frac{Q(E)}{\pi^{3/2} r_{\text{diff}}^3} e^{-r^2/r_{\text{diff}}^2}$ with the effective diffusion distance $r_{\text{diff}}^2 = 4Dt$. The CR intensity is then given by $I = (c/4\pi)n$.

This approach assumes the instantaneous injection of CRs by the SN. In reality, the injection is delayed until the remnant expands far enough to release the CRs. While the release is therefore not actually instantaneous, we take this approximation as reasonable for our cases of interest.

SN total energies range from 0.5 to 4.0 $\times 10^{51}$ erg (Kasen & Woosley 2009), with a conversion efficiency to a CR total energy of about 10% (Dermer & Powale 2013; Cristofari et al. 2021). Here we assume an SN total energy of $2.5 \times 10^{51}$ erg and a conversion efficiency of 10%, yielding a total CR energy of $2.5 \times 10^{50}$ erg. As noted in Melott et al. (2017), this total energy is significantly higher than that used in Gehrels et al. (2003), the only other work reporting detailed modeling of the effects on Earth's atmosphere from nearby SNe. In that work, the SNCR flux is approximated by a simple scale factor applied to the normal background Galactic CR (GCR) flux, rather than explicit modeling of the flux.

As noted in Melott et al. (2017), the magnetic field environment in the intervening space is critical in determining the actual SNCR flux arriving at Earth. If a field line runs directly between the SN and the solar system, the SNCR intensity will be much higher than modeled here. On the other hand, if the predominant field is perpendicular to the line of sight the SNCR flux will be strongly attenuated. We consider the diffusive transport case a middle-ground estimate between those two extremes.

### 2.1. SNCR Flux Results

In Figure 1 we show the calculated CR flux spectrum (times $E^2$, so that the area under the curve is proportional to the total energy between limits) for all four distance cases (100, 50, 20, and 10 pc), at several times. Times are measured from the arrival of the first photons from the SN. Also shown for reference in each case is the current GCR background.

Panel A in Figure 1 (the 100 pc case) compares to Case C in Figure 1 of Thomas et al. (2016), and panel B in Figure 1 (the 50 pc case) compares to Case B in Melott et al. (2017). In those past works the flux was overestimated at energies below about $10^5$ GeV and times earlier than about 3000 yr (100 pc case) and energies below about $10^4$ GeV and times earlier than about 1000 yr (50 pc case). The main consequences of that difference, for our purposes, is the difference in the calculated atmospheric ionization caused by the CR particles, and the difference in the sea-level secondary muon flux. Panels C and D are new cases, with SNe at 20 and 10 pc.

## 3. Atmospheric Ionization and Chemistry Modeling

Atmospheric ionization by high-energy photons and CRs has important effects on chemistry. Odd-nitrogen oxide compounds (NO$_y$, most importantly NO and NO$_2$) are produced, leading to the destruction of stratospheric ozone and subsequent rainout of HNO$_3$ (Thomas et al. 2005). Ionization by SNCRs in this work was computed using the tables from Atri et al. (2010). Those tables provide the ionization rate (ions cm$^{-2}$ s$^{-1}$) as a function of altitude for different primary proton energies from 300 MeV to 1 PeV, at 46 altitude bins, from the ground to 90 km. Here we used primary proton energies between 10 GeV and 1 PeV.

The effects on atmospheric chemistry by SNCR ionization were simulated using the NASA GSFC 2D coupled chemistry-radiation-dynamics model. This model is an updated version of that used in Melott et al. (2017) and Thomas et al. (2016). In particular, the version used here includes feedback on temperature fields due to changes in constituents such as ozone, which was not a feature of the previous version. This allows for more accurate simulation of the chemistry features as well as providing information about the associated temperature changes, which we were not able to evaluate previously. In addition, the spatial resolution of the model has been increased to 4° latitude by 1 km altitude (from 10° latitude by 2 km altitude in the previous version), from the ground to about 92 km. A detailed description of this version of the model can be found in Fleming et al. (2011), see also Fleming et al. (2015) and Airapetian et al. (2017). We run the model in a preindustrial state, with anthropogenic compounds (such as CFCs) set to zero.

Ionization rates as a function of altitude for a particular case are generated by convolving the SNCR flux (essentially the number of incident protons at each energy) with the tables from Atri et al. (2010). The results are translated into the production of NO$_y$, assuming 1.25 NO$_y$ molecules per ion pair (Porter et al. 1976) at all altitude levels. The NO$_y$ production rate is read in to the atmospheric model and the model runs for 20 yr until steady-state conditions were reached, simulating a steady-state flux of CRs. We then examine the concentrations of various constituents, especially O$_3$, comparing a run with SNCR ionization input to a control run without that ionization source.

We use the updated version of the GSFC model with the revised SNCR ionization profiles to revisit the atmospheric response in the 100 and 50 pc cases. In addition, we have performed new modeling of the 20 pc case.

We also sought to model the case of a 10 pc SN, roughly the distance commonly used as the standard "lethal" distance, based on Gehrels et al. (2003), who used a simple scaling up of the standard GCR ionization to simulate the effects of SNCRs. Here, we use the CR flux calculated as described above (shown in Figure 1). As may be seen in Figure 2 (panel D), the induced ionization for the 10 pc case is nearly three orders of magnitude above the background GCR ionization. When used in the newer GSFC model this very large ionization input produced unphysical results, due to numerical overflows. Therefore, in order to model the 10 pc case we have utilized the older, simpler version of the GSFC model (Thomas et al. 2005), which was able to handle the large ionization input and produced reasonable results, as compared to the other cases. We discuss the results of all cases below.

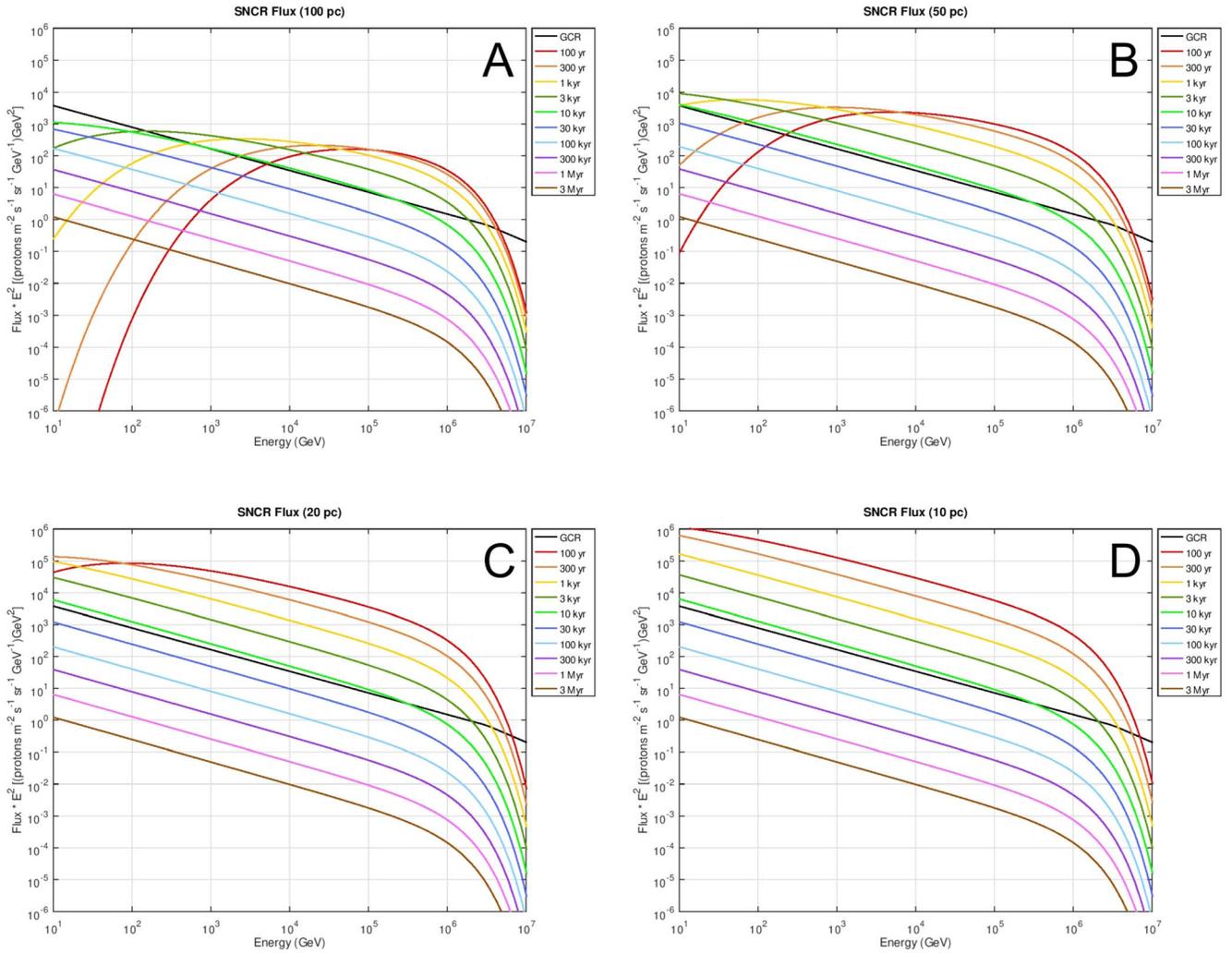

**Figure 1.** CR flux spectrum (times $E^2$) for an SN at 100 pc (A), 50 pc (B), 20 pc (C), and 10 pc (D), at times from 100 yr to 3 million years after the arrival of the first photons from the SN. The present-day GCR background flux spectrum is shown in each panel (black line) for reference.

### 3.1. Atmospheric Ionization Results

In Figure 2 we show atmospheric ionization rate profiles computed using the fluxes shown in Figure 1, for the same distance cases and times.

Comparing to previous work, here in the 100 pc case (Figure 2(A)) we find significantly smaller ionization than that in Thomas et al. (2016), due to the strong suppression at lower energies and earlier times not included in the previous work. Recall that the results presented here were generated using the assumptions of Case C in Thomas et al. (2016) and Case B in Melott et al. (2017) and all comparisons between this work and previous results are for that case. In Thomas et al. (2016), for the 100 pc case, ionization is maximized at the 500 yr mark, with a factor of a few enhancement over the background GCR-induced ionization. Here we find that there is only a small enhancement over the GCR ionization and only at the lowest altitudes.

There is less of a difference in the 50 pc case (Figure 2(B)), comparing the results here to those in Melott et al. (2017). In the previous work a maximum enhancement over background of about 50 times was seen, while here the increase is closer to 10 times. However, the timing of that increase changes from the 100 to 300 yr range in the previous work to the 300–1000 yr range in this work. Essentially, the time of maximum impact is shifted later due to the early suppression included here, which was neglected in the previous work.

In Figure 2 panels C and D we present the ionization results for the 20 and 10 pc cases, respectively. As may be expected, the ionization rates are higher in these cases and the time variation is simpler due to the less-pronounced (or absent) early suppression. Maximum enhancement over the background in the 20 pc case is around 200 times and in the 10 pc case around 1000 times.

### 3.2. Atmospheric Chemistry Results

In order to evaluate the impact on $O_3$, we computed the percent difference in the column density values between the SNCR cases and a control run of the model with no additional ionization input. In Figure 3 we show the globally averaged percent difference in the column density of $O_3$ for the 20, 50, and 100 pc cases at the times of maximum ionization.

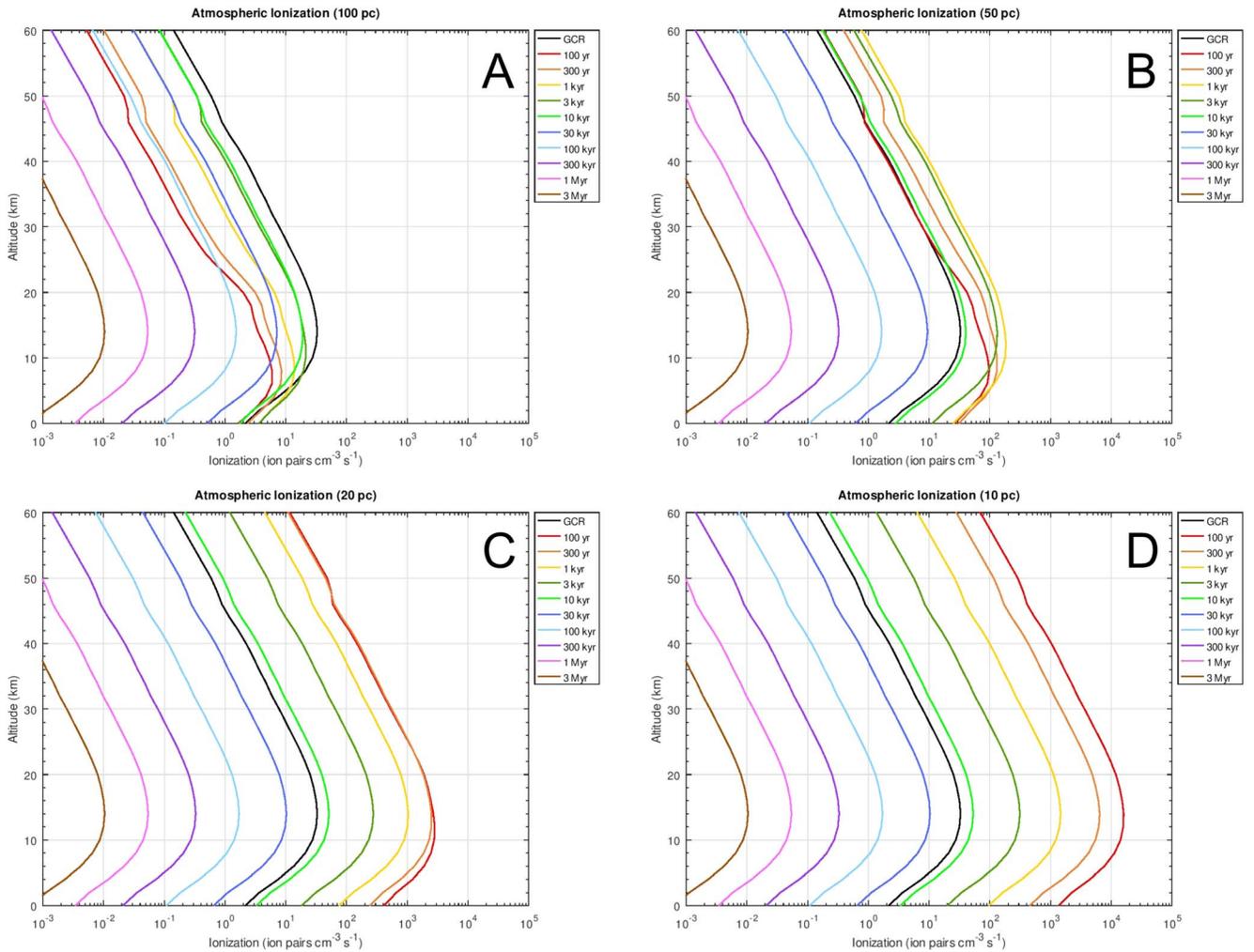

**Figure 2.** Atmospheric ionization rate profiles for ionization caused by CRs from an SN at 100 pc (A), 50 pc (B), 20 pc (C), and 10 pc (D), at times from 100 yr to 3 million years after the arrival of the first photons from the SN. The present-day ionization profile caused by GCRs is shown in each panel (black line) for reference.

For the 100 pc case the maximum ionization occurs at the 3000 yr mark. Using those ionization results in the atmospheric chemistry model we find a globally averaged depletion around 2%. This is slightly smaller than that reported in Thomas et al. (2016), which is expected due to the lower ionization in this work compared to that past work.

For the 50 pc case the maximum ionization occurs at the 1000 yr mark. In this case we find a globally averaged depletion around 17%. This is somewhat smaller than the 25% reported in Melott et al. (2017), which is again attributable to the smaller ionization values in this work compared to that past work.

For the 20 pc case (new to this work) the maximum ionization occurs at the earliest modeled time (100 yr). The resulting globally averaged depletion is about 62%. In Figure 4 we show the percent difference (compared to the control) in the $O_3$ column density as a function of latitude and time for this case. Notice that there is significant depletion, over 30%, at all latitudes, and up to 87% in the polar regions.

As discussed above, for the 10 pc case we have used the simpler GSFC model version (without temperature feedback and lower spatial resolution) since the newer, more complex model is not able to simulate conditions under the very high ionization rates for this case. Using that simpler model we find a globally averaged depletion only slightly higher than the 20 pc case. This is not surprising given that the $O_3$ depletion scales roughly with energy input as a power law with index ∼0.3 (Thomas et al. 2005), reflecting the fact that the depletion "saturates" as it approaches 100%. In Figure 5 we show the percent difference (compared to the control) in the $O_3$ column density as a function of latitude and time for this case. Note that in this case there are isolated areas in the polar regions where the $O_3$ column density is briefly higher than the control case. This effect was explored for the case of a gamma-ray burst (GRB; Thomas et al. 2005) and is the result of an extremely high production of nitrogen oxides which, through photochemical processes, result in the short-term production of $O_3$ at the start of south polar spring.

Finally, we investigated the possible effects on atmospheric chemistry under increased lightning rates, as suggested in Melott & Thomas (2019). The GSFC model includes lightning as an input to nitrogen oxide production, which near Earth's surface participates in chemistry that increases the $O_3$ concentration. We modified the model to scale up the lightning rates based on the enhancement of ionization in SNCR cases over the background GCR ionization. Given that little is known about the magnitude of change in lightning for a given change in ionization, we used a simple linear scaling. Past work examined the effects of increased surface-level $O_3$ in the case of a GRB (Thomas & Goracke 2016), finding that the concentration was nearly doubled for a particular case, but the increase was not large enough to cause biological damage. In the 20 pc case considered here we find a maximum increase over the control of about 2.5%. Since this is much smaller than the GRB case we will not consider it further.

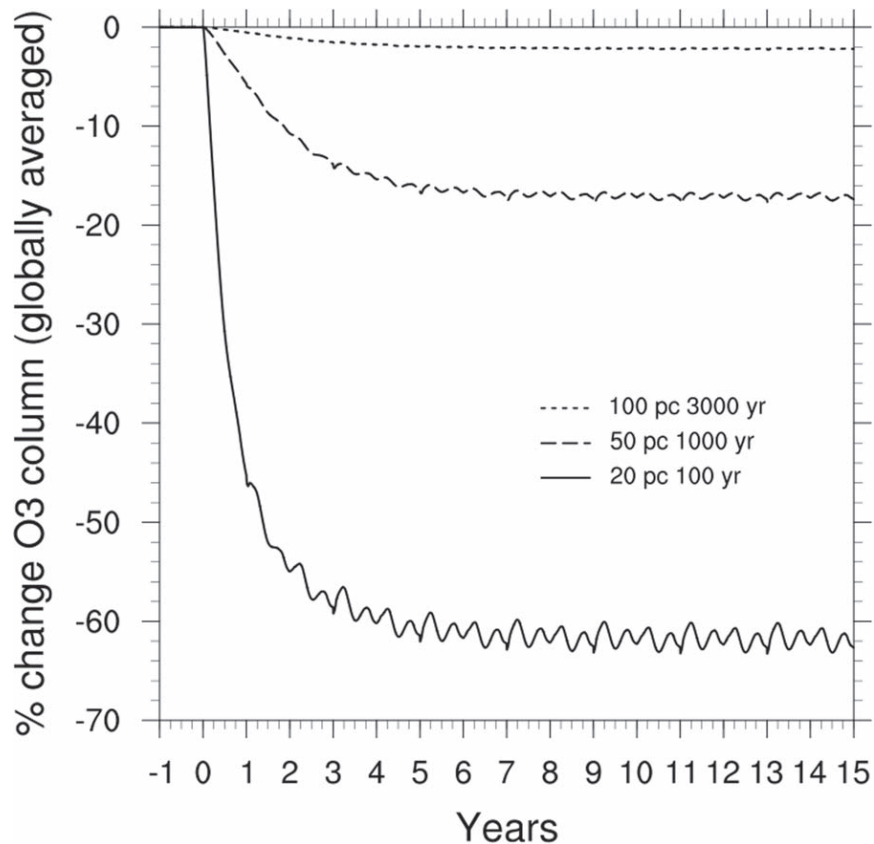

**Figure 3.** Globally averaged change (percent difference vs. control) in the atmospheric $O_3$ column density, assuming steady-state irradiation after time 0 for an SN at 100 pc (dotted line), 50 pc (dashed line), and 20 pc (solid line). The times (3000, 1000, and 100 yr) correspond to those in Figures 1 and 2, and represent the epoch of maximum $O_3$ depletion for each case.

### 4. Sea-level Muon Flux

Following Thomas et al. (2016) and Melott et al. (2017) we calculated the sea-level muon flux by convolving our calculated SNCR spectra with the table of Atri & Melott (2011) and then used the result to compute the radiation dose. As discussed in Melott et al. (2017), radiation doses for muons have not been measured for the energies relevant to this work, so we have used the energy deposition in water and a radiation-weighting factor of 1 (the same as for electrons).

In Figure 6 we show the results of the muon flux calculations. Comparing our results for the 50 pc diffusive case considered here (Figure 6, panel A) to the equivalent 50 pc case (Case B) in Melott et al. (2017), the maximum flux occurs at the 1000 yr (instead of 100 yr) mark and the maximum is roughly an order of magnitude smaller than seen in the previous work. This is due to the lower-energy CR suppression included here that was previously neglected. In fact, the Melott et al. (2017) results more closely match our calculations for the 20 pc case (Figure 6, panel B).

In Table 1 we present the calculated annual muon radiation dose, in mSv, for our 20 and 50 pc cases, and those from Case B in Melott et al. (2017) for comparison.

Unlike the $O_3$ depletion results discussed above, where this updated study finds roughly similar results to the past works, in the case of the muon dose there is a significant difference, with previous works overestimating the dose in the 50 pc case by roughly an order of magnitude. This difference can be attributed to the strong suppression previously neglected in Melott et al. (2017) at the energies relevant for muon flux (at the early-time points).

### 5. Discussion and Conclusions

#### 5.1. Updates to Terrestrial Effects of Nearby SNe

The goals of this work were twofold. First, to reevaluate recent works (Thomas et al. 2016; Melott et al. 2017) on nearby SNe at 100 and 50 pc in the diffusive transport CR case, since those works neglected the early-time suppression at lower CR energies. Inclusion of this suppression leads to lower overall CR fluxes at early times, lower atmospheric ionizations, smaller resulting ozone depletion, and lower sea-level muon radiation doses. The differences are most pronounced for the 100 pc case with only a small enhancement in low-altitude ionization and a small globally averaged decrease in column $O_3$. We can conclude that, for the specific parameters

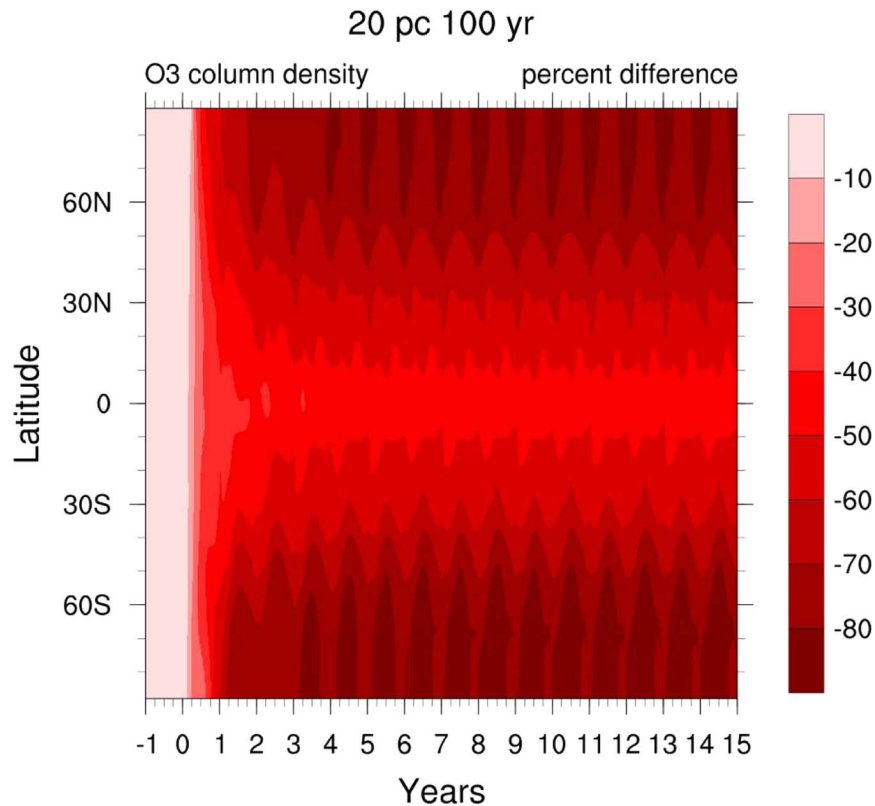

**Figure 4.** Percent difference (vs. control) in the $O_3$ column density, as a function of latitude and time for the 20 pc SN case at 100 yr after the arrival of the photons, assuming steady-state irradiation after time 0.

considered in this study, 100 pc is too far to result in noticeable terrestrial effects.

For the 50 pc case, the enhancement in low-altitude ionization and the globally averaged decrease in column $O_3$ are smaller than those reported in Melott et al. (2017), but the differences are relatively small and do not significantly change the conclusions of that work or subsequent works based on those results (Thomas 2018; Melott et al. 2019; Melott & Thomas 2019; Melott et al. 2020; Thomas & Ratterman 2020). We expect that a 50 pc SN would have noticeable terrestrial effects, but perhaps not dramatic enough to be obvious in the fossil record.

Our second goal was to perform new modeling for SNe at 20 and 10 pc. While geochemical evidence supports the occurrence of at least one SN within the 50–100 pc range about 2.6 Mya, there is no direct evidence of SNe closer than 50 pc. However, it is likely that there have been at least 1–2 SNe in the 10–20 pc range within the last 500 Mya (Melott & Thomas 2011).

We find that already at 20 pc the SNCRs cause major global $O_3$ depletion, which would last on the order of centuries to millennia. The maximum $O_3$ depletion is limited by essentially running out of $O_3$ to destroy, coupled with $O_3$ production mechanisms that offset depletion and continue regardless of the ionization input, so that distances closer than 20 pc result in only marginally greater depletion.

Regarding the sea-level muons, we found that Melott et al. (2017) dramatically overestimated the radiation dose at 50 pc. In fact, those past values are more comparable to our results for a 20 pc SN. Based on these results, it appears unlikely that muon radiation is a significant threat to the biosphere for SNe beyond 20 pc. This finding implies that the results reported in Melott et al. (2019) are likely not accurate for a 50 pc SN, but would apply for an SN at 20 pc.

It should be noted that the present study includes only the effects of CRs associated with SN explosions. These changes occur centuries and longer after any electromagnetic emission is received. Recent work (Brunton et al. 2023) indicates that the effects of SN X-ray emission may be as important as CR impacts. However, the two are not coincident in time, with X-ray emission coming in the months to years after the initial outburst and the CR flux becoming significant later. High-energy photons (i.e., X-rays) will have similar atmospheric effects, but that impact will only persist for a decade or so (Gehrels et al. 2003; Thomas et al. 2005), while the CR impact will persist for 100–1000s of years.

### 5.2. Implications for the "Lethal" SN Distance

Perhaps most importantly, this study indicates that the "lethal" SN distance should be closer to 20 pc rather than the typically quoted 8–10 pc. With a larger distance, such SN events become more common and may be associated with more instances of extinctions, and possibly climate change periods, during the Phanerozoic.

Given the uncertainties in the total SN energy, efficiency of conversion to CR energy, and CR transport, it is worth considering the potential extreme range in the lethal distance. The total SN energy has a possible range from 0.5 to $4 \times 10^{51}$ erg (Kasen & Woosley 2009); therefore the SN energy could be smaller than our assumption by a factor of 0.2, or larger by a factor of 1.6. The efficiency of conversion to CR total energy is probably about 10%, but may be as high as 30% (Dermer & Powale 2013; Cristofari et al. 2021); therefore the total CR energy could be larger than our assumption by a factor of three. Finally, the efficiency of transport of CRs to Earth can vary widely, depending primarily on the magnetic field environment of the intervening space. A minimum transport case would be the blocking field of Case B in Thomas et al. (2016). A maximum transport case would be the connecting field line of Case A in Melott et al. (2017). To allow for a simple comparison, let us consider the CR flux at $10^6$ GeV. For a 100 pc SN, the flux at $10^6$ GeV in Thomas et al. (2016) Case B is about 30 times smaller than the corresponding flux in this work. For a 50 pc SN, the flux at $10^6$ GeV in Melott et al. (2017) Case A is about 300 times larger than the corresponding flux in this work.

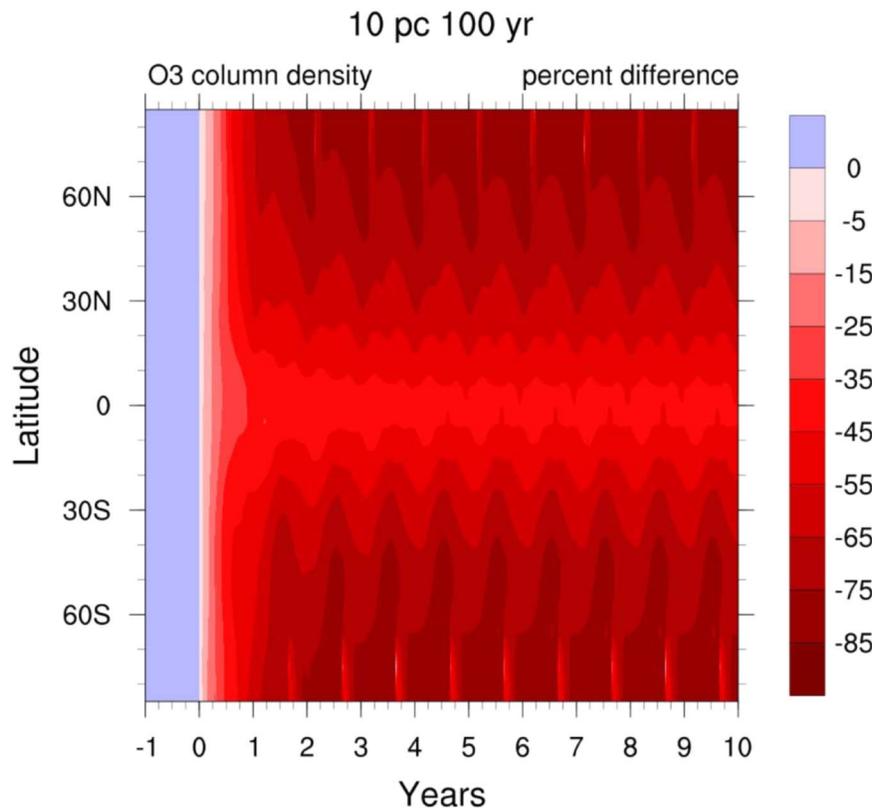

**Figure 5.** Percent difference (vs. control) in the $O_3$ column density, as a function of latitude and time for the 10 pc SN case at 100 yr after the arrival of the photons, assuming steady-state irradiation after time 0.

Combining the increasing factors (1.6 times the total energy, 3 times the conversion efficiency, and 300 times the transport efficiency), the received flux could be as much as 1400 times larger than our present results. Conversely, combining the decreasing factors (0.2 times the total energy and 1/30 times the transport efficiency), the received flux could be as small as 0.007 times our present result.

Translating these factors into modifications of the lethal distance requires connecting the CR flux to ozone depletion. In Thomas et al. (2005) it was found for the case of photons from a GRB that the globally averaged ozone depletion scales roughly as the total received fluence to the 0.3 power. Since the depletion mechanism is essentially the same for CRs as it is for high-energy photons, we will assume that relation holds here as well. Therefore, in the maximum case, an increase in flux of 1400 times could mean a lethal distance increase by as much as eight times, from 20 to 160 pc. In the minimum case, a decrease in flux by a factor 0.007 could mean a lethal distance change factor of about 0.2, from 20 pc to about 4 pc.

Overall, then, given the range of possible SN total energies, conversion efficiency to CR, and variation in transport, the lethal distance may lie anywhere between 4 and 160 pc. The most important, and most variable, factor here is that of CR transport. The interstellar magnetic field environment around a star system has a dramatic effect on the received CR flux from any nearby SN. Each factor involved is subject to uncertainty and in particular the maximum case represents a possible but very special configuration. The case presented in this work, we believe, represents a realistic middle ground and therefore we consider 20 pc a good working value for the lethal distance. Again, we emphasize that this work considers only the effect of CRs, not of any high-energy photons that may be associated with the prompt or later phases.

It should also be noted that determining the "lethality" of an astrophysical ionizing radiation event based on the globally averaged $O_3$ depletion level is highly uncertain and more work is needed to establish the connection between the specific $O_3$ depletion level and the magnitude of the impact on the biosphere.

A 20 pc lethal distance is particularly significant since it has been suggested (Fields et al. 2020) that extinctions at the end of the Devonian period may have been triggered by an SN explosion at roughly 20 pc. In that work it was assumed that 20 pc was somewhat beyond the "lethal" distance necessary to initiate a mass extinction. Our results here indicate that in fact 20 pc may well be a mass-extinction-inducing distance. The less-severe biodiversity crisis at the end of the Devonian may indicate a larger distance of any associated SN, perhaps as far

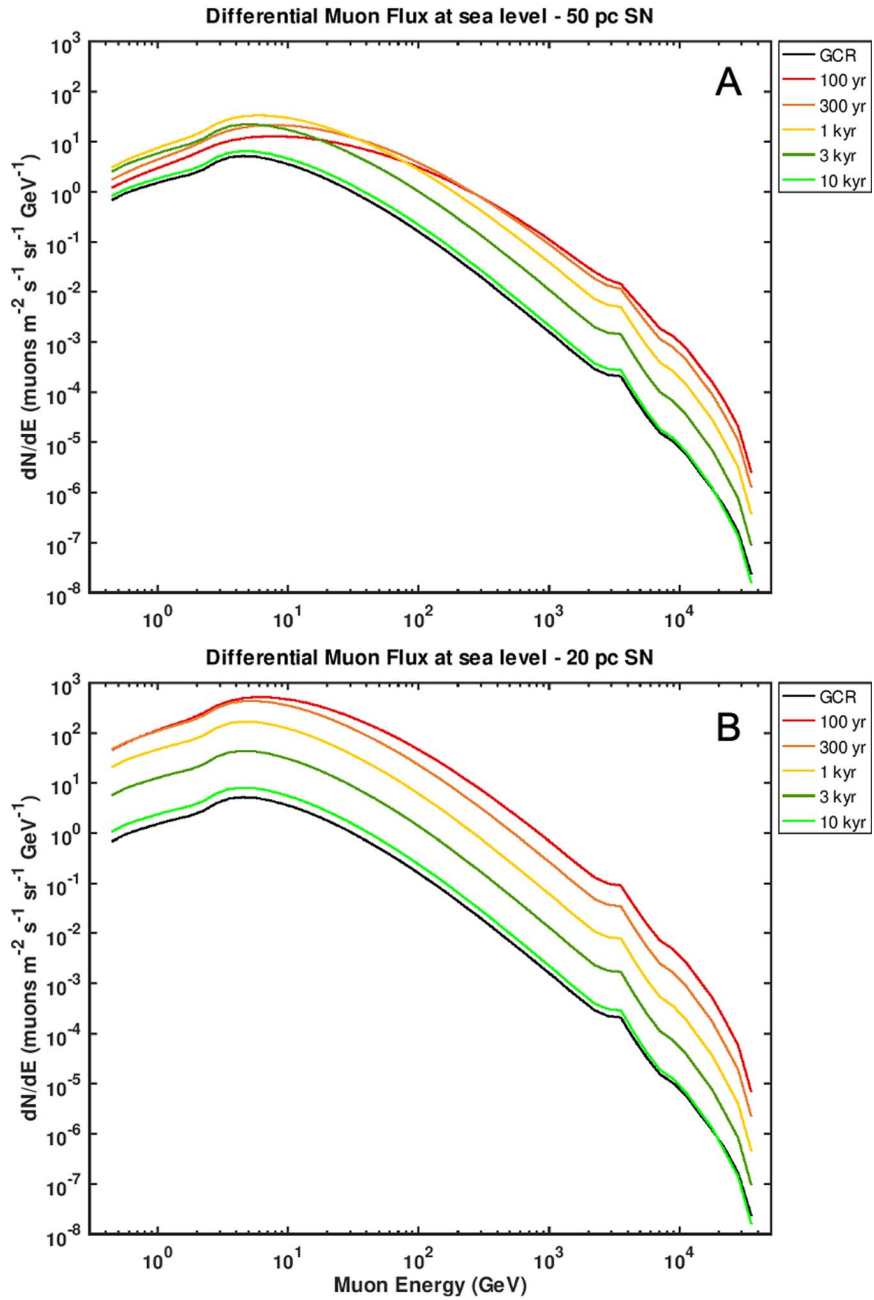

**Figure 6.** Differential muon spectra at sea level for an SN at 50 pc (A) and 20 pc (B) at several times (corresponding to those in Figure 1). The black line represents the typical present spectrum, primarily as a result of showers initiated by GCRs.

**Table 1**
Annual Ground Level Muon Radiation Dose, in Units of mSv, for Two SN Cases in This Work and for Case B in Melott et al. (2017)

| Time   | 20 pc | 50 pc | 2017 50 pc Case B |
|--------|-------|-------|-------------------|
| 100 yr | 36.9  | 2.66  | 31.5              |
| 300 yr | 20.9  | 2.87  | 21.1              |
| 1 kyr  | 6.28  | 2.22  | 8.36              |
| 3 kyr  | 1.49  | 0.966 | 2.61              |
| 10 kyr | 0.266 | 0.23  | 0.587             |

**Note.** For comparison, the present-day average annual dose on the ground due to muons is about 0.20 mSv.

as 50 pc. Importantly, a larger distance increases the likelihood of an SN event, thereby potentially strengthening the argument presented in Fields et al. (2020).

In addition to the long-term CR effects studied here, recent work (Brunton et al. 2023) indicates that the effects of SN X-ray emission may be as important as CR impacts, though much shorter lived, even out to 50 pc. SNe as far as 50 pc may therefore present both a short- and long-lived lethal threat for Earth and any Earth-like planet with similar atmospheric conditions. This greatly increases the likelihood that at least one and perhaps several extinctions in the fossil record may be associated with nearby SNe.

The authors thank M. Kachelrieß and A. Melott for helpful discussions regarding the updated SNCR modeling and this manuscript, and the anonymous reviewer whose comments greatly improved the manuscript. B.T. acknowledges publication cost support from the Washburn University Office of the Vice President for Academic Affairs.

## Data Availability

The data underlying this article, as well as code used to calculate the SNCR flux, atmospheric ionization rates, and muon flux, are freely available at https://zenodo.org/record/7764996 . The NASA GSFC 2-D coupled chemistry-radiation-dynamics model version used in this study can be accessed at https://github.com/brianthomas-washburn/GSFC2DCoupledAtmo_Supernova.

## ORCID iDs

Brian C. Thomas  https://orcid.org/0000-0001-9091-0830
Alexander M. Yelland  https://orcid.org/0000-0002-1462-0265